\documentclass[fleqn,usenatbib]{mnras}

\usepackage{newtxtext,newtxmath}
\usepackage{appendix}

\DeclareRobustCommand{\VAN}[3]{#2}
\let\VANthebibliography\thebibliography
\def\thebibliography{\DeclareRobustCommand{\VAN}[3]{##3}\VANthebibliography}

\usepackage{graphicx}	
\usepackage{tabularx}
\usepackage{amsmath}	

\title[viewing angle of GRB 221009A]{Determining the viewing angle from TeV light curve of GRB 221009A}

\author[L. Zhou and Y.-C. Zou]{
Lin Zhou$^{1}$,Yuan-Chuan Zou$^{1,2}$\thanks{E-mail: zouyc@hust.edu.cn (Y-CZ)}
\\
$^{1}$Department of Astronomy, School of Physics, Huazhong University of Science and Technology, Wuhan 430074, China\\
$^{2}$Purple Mountain Observatory, Chinese Academy of Sciences, Nanjing, 210023, China\\
}

\date{Accepted XXX. Received YYY; in original form ZZZ}

\pubyear{2015}

\begin{document}
\label{firstpage}
\pagerange{\pageref{firstpage}--\pageref{lastpage}}
\maketitle

\begin{abstract}
Gamma-ray bursts (GRBs) are among the most powerful explosive events in the universe. LHAASO recently observed the most luminous one: GRB 221009A, and unveiled its TeV light curve.  The light curve exhibits a distinct jet break at around 670 seconds, enabling the derivation of the viewing angle based on the smoothness of the jet break. We constructed two models with or without considering the high-latitude radiation, where the viewing angle was treated as a free parameter, to fit the TeV light curve. We obtained the viewing angles being 9.4 $\times 10^{-4}$ radians and 5.9 $\times 10^{-3}$ radians, respectively. These values closely resemble an on-axis scenario, given the opening angle is 1.4 $\times 10^{-2}$ radians.

\end{abstract}




\section{Introduction}\label{sec:intro}

Gamma-ray bursts (GRBs) are one of the most powerful explosive events in the universe, releasing extremely highly energetic gamma-ray radiation. For a long time, GRBs have been a fascinating research topic for astronomers because their origins and nature involve a series of complex physical processes.

On October 9, 2022, the brightest GRB (GRB 221009A) initially triggered at 13:16:59.988 UTC by the \textit{Fermi} Gamma-Ray Burst Monitor (\textit{Fermi}-GBM; \cite{2009ApJ...702..791M}) in the gamma-ray band \citep{2022GCN.32636....1V,2022GCN.32642....1L}. This event was also detected \citep{2022GCN.32658....1P} by the \textit{Fermi} Large Area Telescope (\textit{Fermi}-LAT; \cite{2009ApJ...697.1071A}). The Large High Altitude Air Shower Observatory (LHAASO; \cite{2019arXiv190502773C}) also reported the detection of gamma rays, reaching up to 18 TeV \citep{2022GCN.32677....1H}. \textit{Swift} Burst Alert Telescope (\textit{Swift}-BAT; \cite{2005SSRv..120..143B}) also detected the same event approximately an hour after the initial trigger. About 143 seconds later, the \textit{Swift} X-ray Telescope (\textit{Swift}-XRT; \cite{2005SSRv..120..165B}) pinpointed the target, with the \textit{Swift} Ultraviolet/Optical Telescope (UVOT; \cite{2005SSRv..120...95R}), locating it at (RA, DEC) = (288.26452 $^\circ$, 19.77350 $^\circ$) with a 90\% confidence error radius of about 0.61 arcsec \citep{2022ATel15650....1D}. The event was quickly classified as a gamma-ray burst \citep{2022GCN.32636....1V}, and according to X-shooter/VLT reports. it occurred at a redshift of 0.151 \citep{2022GCN.32765....1I}, confirming it as the brightest event ever recorded by any gamma-ray burst detector. After two days, reports of a soft X-ray ring due to dust scattering were released, based on observations from the \textit{Swift}-XRT \citep{2022GCN.32680....1T}. Multiwavelength observations were reported later\citep[e.g.][]{2022GCN.33002....1K,2022GCN.32653....1B}. 

The variation in the light curve and the corresponding physical processes of GRB 221009A have become a highly discussed research topic. The half-opening angle and the viewing angle are elements that influence the light curve during the afterglow decay phase. 
The half-opening angle $\theta_{\rm jet}$ is the angle at which the cone-shaped region extends from the central axis of the jet towards one side. The viewing angle $\theta_{\rm obs}$, also known as the observer's angle, 
represents the angular deviation of the observer from the cone axis. Both lead to a steepening decay in the late afterglow. 

GRB 221009A is believed to have a narrow opening angle and a slightly off-axis direction, as evidenced by the jet break in the later stages of the afterglow light curve. 

Quite amount of studies have been devoted to the investigation of the jet opening angles of this Gamma-Ray Burst \citep[e.g.][]{2023ApJ...947...53R,2023ApJ...946L..24W,2023ApJ...946L..23L}. However, limited attention has been given to the consideration of off-axis angles.
\cite{2023Sci...380.1390L} estimated a half-opening angle $\theta_{\rm jet}$ to be approximately $0.8^\circ$, equivalent to 0.140 radians. This determination was made using the jet break time and applying their late afterglow model \citep[e.g.][]{1999ApJ...519L..17S}, assuming a viewing angle of zero. \cite{2023SciA....9I1405O}'s  findings yielded $\theta_{\rm obs} \lesssim 0.016$ rad and $\theta_{\rm jet} \ge $ 0.4 rad with a structured jet model. This study discussed a situation about a structured jet with extended wings and determined the viewing angle.

\cite{2023MNRAS.524L..78G} built a semi-analytical model to analyze an adiabatic spherical blast wave and extended the model to angular structured jets. They found that the viewing angle was comparable to the core angle of the energy profile, $\theta_{\rm obs} \sim \theta_{\rm c,\epsilon}$, which was determined to be 0.02 rad. This was employed to explain the high fluence of the gamma-ray emission and to interpret the multi-waveband light curve.

Polarimetry was used to infer the viewing angle \citep[e.g.][]{1999MNRAS.309L...7G}. \cite{2023ApJ...946L..21N} took a set of parameters to give a polarization consistent with the IXPE spectropolarimetric measurement and found it favors a jet opening angle approximately
$1.5^\circ$, and a viewing angle about 2/3 of the jet opening angle, which correspond to 0.026 rad and 0.017 rad, respectively.

\begin{figure}
    \centering
    \includegraphics[width=1\linewidth]{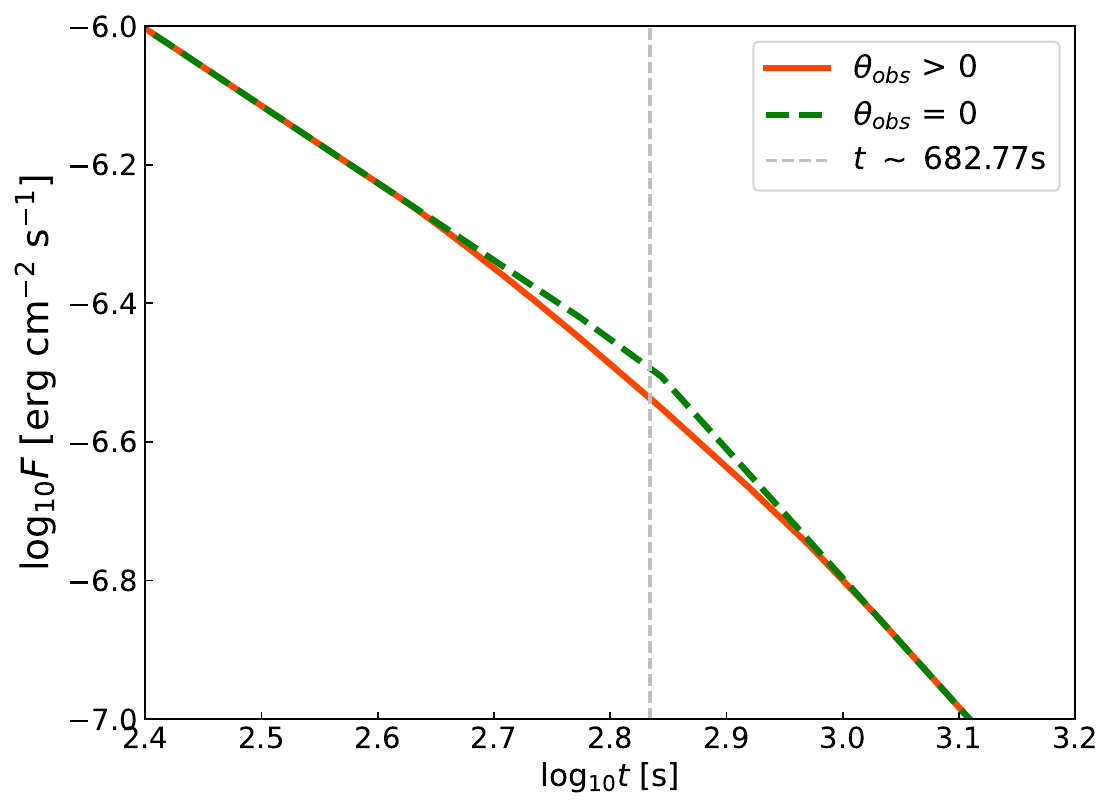}
    \caption{Illustrating the modification of fitting results for the GRB afterglow jet break through incorporation of the viewing angle effect. The green dashed line is the fitting without viewing angle ($\theta_{\rm obs} =0$) while the red solid line accounts for the viewing angle effect. Gray dashed line represents jet break time.}
    \label{fig:show-obs}
\end{figure}

In this work, we focus on the TeV light curve around the jet break. 
We draw Fig. \ref{fig:show-obs} as a sketch. If the viewing angle is zero, one may expect a sharp jet break once the emission of the jet edge theoretically reaches the observer. If the viewing angle is non-zero, the different part of the edge is then gradually reaches the observer, and consequently the sharp break is smoothed.
By fitting the light curve with a top-hat jet model, we aim to determine the viewing angle.
We present the details of the model and the results in section \ref{subsec:without} and \ref{subsec:with}. Data selecting is shown in section \ref{subsec:data} and the method of fitting is presented in section \ref{sec:MCMC}. The results are shown in section \ref{subsec:result}. The conclusions and discussions are shown in section \ref{sec:con}.

\section{Constraining the viewing angle}\label{sec:model}
\subsection{top-hat model without considering high-latitude emission} \label{subsec:without}

After the peak of GRB afterglow light curve, the energy flux follows a power-law decay with an index $\alpha$, which can be influenced by the visible area and the viewing angle $\theta_{\rm obs}$. With the viewing angle, the power-law decay is smoothed. Following the study by \cite{2023Sci...380.1390L}, which assumes no high-latitude radiation, we constructed a simple geometry model based on the top-hat model. This model describes the visible area, which varies with time and the viewing angle. 

\begin{figure}
    \centering
    \includegraphics[width=1\linewidth]{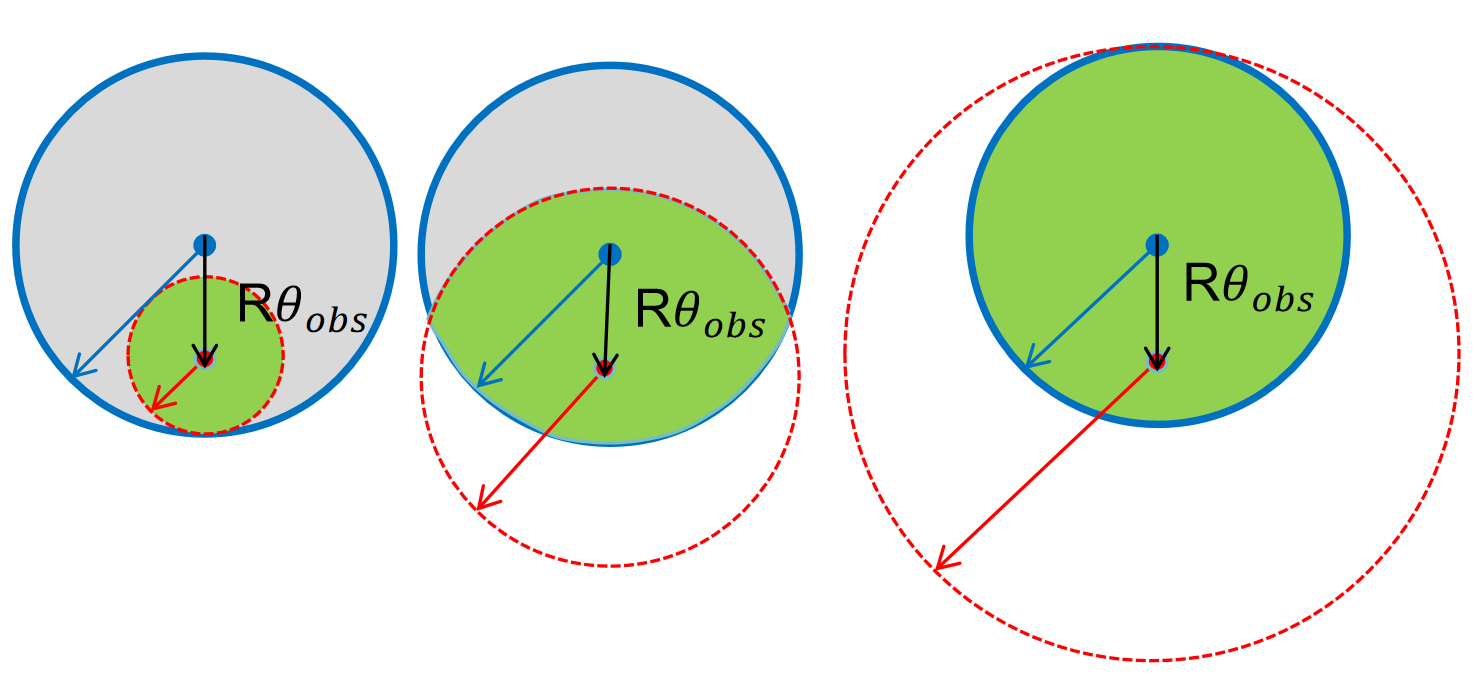}
    \caption{Diagram of the variation in jet area as observed by an observer. \textit{Left}: when $t = t_{\rm in}$, the observing area is first tangent to the jet edge; \textit{ Mid}: when $t_{\rm in} < t < t_{\rm out}$, the observing circle overlaps with a part of the jet; \textit{Right}: when $t = t_{\rm out}$, the observing area covers the jet and is tangent to the edge for the second time. }
    \label{fig:jet-break}
\end{figure}

At a distance $R$, as shown in Fig. \ref{fig:jet-break}, with time passing and Lorentz factor $\Gamma$ decreasing, the visible area ($R \theta$, green part) expands. The blue circle represents the extent of the jet, with the blue dot in the center indicating the direction of the jet. The dashed red circle represents the observable range for the observer as the Lorentz factor decreases.
The green region represents the portion of the jet that actually enters the observer's observable range, and the gray area represents the part of the jet temporarily invisible to the observer. $t_{\rm in}$ and $t_{\rm out}$ represent the moments when the observing area first and second tangentially touch the edge of the jet, respectively.

This simple geometry model constructs a simple physical scenario, as exemplified by \cite{2001Natur.414..853W}. 
Because of the relativistic beaming effect, the radiation outside the angle $1/\Gamma$ is ignored, while the radiation inside is considered as viewing angle independent. It is denoted as Model 1 or geometric model. 
(A more realistic model by considering the precise frame transformation is discussed in section \ref{subsec:with}, which is denoted as Model 2.)
As the Lorentz factor decreases, the visible area expands, taking on the form of a circle. For $t < t_{\rm in}$, the visible area is in circular shape, as shown in the left panel of Fig. \ref{fig:jet-break}. During this period, the decay index $\alpha$ of the GRB afterglow remains constant. At $t = t_{\rm in}$, which indicates that the edge of the jet first becomes visible to the observer, the afterglow undergoes a steepening decay. The alteration in the decay rate between $t_{\rm in}$ and $t_{\rm out}$ is exclusively determined by the viewing angle and half-opening angle. The latter is proportional to the jet's visible area divided by the total visible area. After $t > t_{\rm out}$, indicating the complete visibility of the entire jet to the observer, the decay rate stabilizes. The difference comparing to the on-axis ($\theta_{\rm obs} =0$) case is during $t_{\rm in} < t < t_{\rm out}$.
Utilizing this simple geometry model, we can derive a plausible value for $\theta_{\rm obs}$.

As relativistic particles emit radiation, their emissions become concentrated within a narrow cone due to the relativistic beaming effect.
This cone has a half angle of $1/\Gamma$, where $\Gamma$ represents the Lorentz factor of the jet. The direction of the cone aligns with the motion of the jet. 
As a result of the relativistic beaming effect, the observer can only detect a portion of the jet that is located within an angular distance smaller than the half angle. In other words, the observable region is confined to a cone-shaped area with a finite opening angle. The half-angle of the narrow cone can be described as
\begin{equation}
    \theta(t) \propto \frac{1}{\Gamma(t)} = \frac{C_1}{\Gamma(t)},
    \label{equ:theta}
\end{equation}
where $C_1$ is a constant. The Lorentz factor $\Gamma(t)$ obeys: 
\begin{equation}
    \Gamma(t) \propto t^{-\alpha} = C_2 t^{-\alpha},
    \label{equ:gama}
\end{equation}
where $C_2$ is another constant determined by the actual value of the Lorentz factor. The power law index $\alpha$ depends on the profile of the number density within the surrounding environment. When the circum-burst number density remains uniformly distributed, the power law index is -3/8 \citep{1998ApJ...497L..17S}. 


Combing equations (\ref{equ:theta}) and (\ref{equ:gama}) derives 
\begin{equation}
    \theta(t)=\cfrac{C_1}{C_2} \cdot t^\alpha.
    \label{equ:thetaA}
\end{equation}
The parameters $C_1$ and $C_2$ are combined into the ratio $C_1/C_2$ in equation (\ref{equ:thetaA}) to represent the $\theta(t)$. 
To determine the variation of $\theta(t)$, we need to consider constants $C_1$ and $C_2$. In this case, we assume that $C_1 = 1$ and $C_2 = \Gamma_{\rm peak} \cdot t_{\rm peak}^{3/8}$.
With equation (\ref{equ:thetaA}), we can estimate the time boundaries $t_{\rm in}$ and $t_{\rm out}$. Once the viewing angle $\theta_{\rm obs}$ is provided, the instances at which the decay rate begins and ceases its steepening is determined. The time boundaries are determined by
\begin{equation}
    \begin{aligned}
    \theta_{\rm jet} - \theta_{\rm obs} &= \frac{1}{\Gamma(t_{\rm in})},\\
    \theta_{\rm jet} + \theta_{\rm obs} &= \frac{1}{\Gamma(t_{\rm out})}.
    \end{aligned}
    \label{equ:border}
\end{equation}

As the external shock decelerates by the swept-up ambient mass, its Lorentz factor decreases with time as $\Gamma \propto t^{-3/8}$. It is reasonable to assume the flux density $F_\nu(t)$ follows 
\begin{equation}
    F_\nu(t) = S_{\rm ratio} \cdot F_{\rm c} t^{-\alpha_1},
    \label{equ:fv}
\end{equation}
where $S_{\rm ratio}$ is a function proportional to the visible area of the jet divided by the total visible area as predicted by theory, $\alpha_1$ is the power law index and $F_{\rm c}$ is an arbitrary constant for fitting. The $S_{\rm ratio}$ satisfies
\begin{eqnarray}
    S_{\rm ratio} = \left \{
    \begin{array}{ll}
         \cfrac{\pi R^2\theta^2}{\pi R^2\theta^2} , & t < t_{\rm in},\\
        \cfrac{S_{\rm overlap}}{\pi R^2\theta^2},   & t_{\rm in} < t < t_{\rm out},        \\
        \cfrac{\pi R^2\theta_{\rm jet}^2}{\pi R^2\theta^2}, & t>t_{\rm out}.
    \end{array}
    \right.
    \label{equ:flux}
\end{eqnarray}
$S_{\rm overlap}$ represents the area where the visible area predicted by theory overlaps with the jet, as shown in the middle panel of Fig. \ref{fig:jet-break}. 
With the law of cosines, $S_{\rm overlap}$ can be expressed by
\begin{equation}
    \begin{aligned}
        S_{\rm overlap} 
             & = R^2 \times \left[2 \arccos \left(\cfrac{\theta_{\rm obs}^2+\theta_{\rm jet}^2-\theta^2}{2\theta_{\rm obs}\theta_{\rm jet}}\right)\cdot \theta_{\rm jet}^2 \right. \\ 
             &+2 \arccos\left(\cfrac{\theta_{\rm obs}^2+\theta^2-\theta_{\rm jet}^2}{2\theta_{\rm obs}\theta}\right) \cdot \theta^2 \\
             &\left. - 2\sqrt{p(p-\theta)(p-\theta_{\rm obs})(p-\theta_{\rm jet})} \right],   
    \end{aligned}
    \label{equ:So}
\end{equation}
where $p = (\theta+\theta_{\rm obs}+\theta_{\rm jet})/2$.

By substituting $S_{\rm overlap}$ into equation (\ref{equ:flux}), we derive an expression for the energy flux based on the previously assumed variables. To simplify calculations, we opt for the logarithmic representation,
\begin{equation}
    \log(F_\nu(t)) = -\alpha_1 \log(t) + b + \log( S_{\rm ratio} ).
    \label{equ:log_f}
\end{equation}
This equation represents the simple model with five parameters: $ \{\theta_{\rm obs}, \theta_{\rm jet}, C_1/C_2, \alpha_1, b \}$, where $b = \rm log(F_{\rm c})$, $C_1$ and $C_2$ are degenerated as one parameter. By combining the TeV light curve data, we estimate the optimal viewing angle. 



\subsection{top-hat model considering high-latitude emission}\label{subsec:with}

Though the geometric model in section \ref{subsec:without} is simple and straightforward, the high-latitude emission may be considered as a comparing model to show whether the simplification is reasonable or not.

Given that the findings of the \cite{2023Sci...380.1390L} were based on the assumption of excluding high-latitude radiation, the previously used half-opening angle $\theta_{\rm jet} = 0.8 ^\circ$ becomes irrelevant. Instead, we should treat $\theta_{\rm jet}$ as a free parameter. Our approach now involves computing the radiation emitted from various segments of the jet and aggregating them to determine the overall energy flux. This requires a detailed analysis of the radiation contributions from different parts of the jet to refine our model and accurately represent the observational data.

The flux from different segments of the jet can be described by the following formula \citep{2002ApJ...570L..61G}:
\begin{equation}
    f_{\nu}(\theta,t) = a^3 f_{\nu/a}(0,at),
\end{equation}
where $a$ satisfies:
\begin{equation}
    a = \frac{1-\beta}{1-\beta \cos(\theta)},
\end{equation}
$\theta$ represents the angular distance between a segment of the jet and the observer, and $\beta$ is velocity.

The flux density of each segment of the jet obeys \citep[e.g.][]{2002ApJ...568..820G}:
\begin{equation}
    f_{\nu}(0, t) \propto t^{-\alpha_2} \nu^{-\beta_2}
\end{equation}
where $\alpha_2$ and $\beta_2$ are parameters that describe how the flux varies with time and frequency. By summing up the flux density of every segment, we can rewrite the flux formula as follows:
\begin{equation}
    F_\nu = F_{\nu, \rm const} t^{-\alpha_2} \nu^{-\beta_2} \iint _{\rm jet} a^{3+\beta_2-\alpha_2}{\rm d}S
\end{equation}
where $F_{\nu, \rm const}$ is a coefficient of the formula.

To compare with the LHAASO observation, which is the flux in the range of 0.3 to 5 TeV, we need to integrate the flux density over frequency to fluence, i.e., $F = \int_{0.3 {\rm TeV}}^{5 {\rm TeV}} F_\nu {\rm d} (\nu/a)$. Then we have
\begin{equation}
    F = \int_{0.3 {\rm TeV}}^{5 {\rm TeV}}   F_{\rm \nu,const}  \nu^{-\beta_2} {\rm d} \nu \cdot t^{-\alpha_2}  \iint_{\rm jet} a^{2+\beta_2-\alpha_2}{\rm d}S,
\end{equation}
where $F_{\rm const} =  \int_{0.3 {\rm TeV}}^{5 {\rm TeV}}   F_{\rm \nu,const}  \nu^{-\beta_2} {\rm d} \nu$ is a parameter for fitting.


By summing up all the energy flux from different angles, we can describe the variation of the flux. This model incorporates several parameters:
$\{F_{\rm const}, C_1/C_2, \alpha_2, \beta_2, \theta_{\rm obs}, \theta_{\rm jet}\}$. When these parameters are provided, the energy flux of the jet is uniquely determined.

\subsection{The data}\label{subsec:data}

LHAASO is a new generation multi-component facility located in Daocheng, Sichuan province of China, at an altitude of 4410 meters. It is designed for the detection of very high energy cosmic rays and gamma rays. The cosmic rays are measured in the energy range between $10^{12}$ and $10^{18}$ eV, while a gamma-ray telescope is operated in the energy range between $10^{11}$ and $10^{15}$ eV. With its capability to measure simultaneously different shower components (electrons, muons, and Cherenkov light), LHAASO enables the exploration of the origin, acceleration, and propagation of CR through its very high-energy spectrum and elemental composition.

GRB 221009A happened to occur within the field of view of LHAASO, at a zenith angle of $28.1^\circ$. During the first 3000 seconds, more than 64,000 photons (above 0.2 TeV) were detected. With a VHE gamma-ray extensive air shower detector, which consists of three interconnected detectors, the LHAASO collaboration analyses the light curve of GRB 221009A and tries to find the best model to explain this extremely intense burst \citep{2023Sci...380.1390L}. In their observation, the TeV photon flux started to increase a few minutes after the GRB trigger and reached its peak approximately 10 seconds later. A decay phase followed, which became more rapid at around 650 seconds after the peak.

They made estimations based on the data collected by LHAASO, which covered an energy range from 0.3 to 5 TeV. With the data collected, they confirmed that the power law break was caused by a jet break.
With a top-hat model, they estimated a half-opening angle of approximately $0.8^\circ$ \citep{2023Sci...380.1390L}.

We use the TeV light curve data of \cite{2023Sci...380.1390L} for the analysis. The energy flux light curve of GRB 221009A exhibits four-segment features: rapid rise, slow rise, slow decay and steep decay. To determine the viewing angle, it suffices to utilize data exclusively from the portions of the light curve associated with slow decay and steep decay. 
To avoid involving the rising part of the light curve, we extract the relevant data commencing approximately 276 seconds after the \textit{Fermi} GBM trigger. This interval also corresponds to around 50 seconds post the TeV trigger.

After a new dataset is done, we employ the \textit{curve\_fit} python package \citep{2020SciPy-NMeth} to determine the optimal initial conditions for the MCMC procedure and start MCMC sampling. 
Since our focus is the viewing angle instead of other parameters, our requirements are confined to the parameters explicitly detailed in Table \ref{tab:paper}. 

\begin{table}
    \centering
    \begin{tabular}{ccccccccccc}
    \hline
        Parameter &$\Gamma_0$ & $t_{\rm peak}$ &$\theta_{\rm jet}$ & $ t_{\rm break}$ & $\alpha_2$ & $\alpha_3$ & $p$\\
    \hline
        value & 560  & 18.0 s  & $0.8^\circ$ & 670 s & -1.115 & -2.21 & 2.2
        \\
    \hline
    \end{tabular}
    \caption{Parameters we choose from \protect\cite{2023Sci...380.1390L}. $\Gamma_0$ is initial Lorentz factor, $t_{\rm peak}$ is the moment of flux peak, $\theta_{\rm jet}$ is the half-opening angle, $t_{\rm break}$ is the moment of the jet break. $\alpha_2$ represents the decay index when light curve started decay, and the $\alpha_3$ is the decay index after the jet break. Notice $\alpha_2$ and $\alpha_3$ here (taken from \protect\cite{2023Sci...380.1390L}) is different from the definition used in our work. $p$ is power law index of the accelerated electron's distribution.} 
    \label{tab:paper}
\end{table}

\subsection{MCMC analysis}\label{sec:MCMC}

The two models have several parameters, which can be collected as two collections: $\Theta_1$ and $\Theta_2$ as
\begin{equation}
    \Theta_1 \equiv \{\theta_{\rm obs},\theta_{\rm jet},C_1/C_2,\alpha_1,b\},
\end{equation}
and
\begin{equation}
    \Theta_2 \equiv \{F_{\rm const},C_1/C_2,\alpha_2,\beta_2,\theta_{\rm obs},\theta_{\rm jet} \}.
\end{equation}

To fit the light curve of GRB 221009A to our model, we utilize a MCMC approach, following \cite{2015ApJ...799....3R}. This algorithm generates samples from the posterior probability distribution $p(\Theta | D)$, which represents the probability distribution of the parameters $\Theta$ given the light curve data. By employing this approach, we gain a better understanding of the overall structure of the posterior distribution and obtain the best estimates for the parameters involved in our model. Ultimately, this method has the potential to provide information regarding the uncertainties associated with the inferred values of individual parameters.

For convenience, we have opted to change some parameters in our model. First, we introduce a new parameter: q, which represents the ratio of $\theta_{\rm obs}/\theta_{\rm jet}$. Second, we set $C_1$ as 1 and $C_2$ as $\Gamma_{\rm peak}t_{\rm peak}^{3/8}$, so that parameter $C_1/C_2$ can be simplified as a single parameter $\Gamma_{\rm peak}$. Third, we simplify $F_{\rm const} $ to $F_0$, which equals log($F_{\rm const}$).

To reduce the models' dimension, for model 1, we fix three parameters: $\theta_{\rm jet}$, $t_{\rm peak}$, p, which can be found in Table \ref{tab:paper}.  The value of $\theta_{\rm jet}$ is set to $0.8^\circ$, which is about 0.0140 rad. $t_{\rm peak}$ is set to 18 s so that the parameter $C_1/C_2$ is determined as $\Gamma_{\rm peak}$ is provided. And $\beta_2$ is set as p-1, which is about 1.2.

By fixing these parameters, we transform the parameter set from $\Theta$ to $\Theta_{\rm fit}$, which will be directly utilized in the MCMC routine, which become
\begin{equation}
    \Theta_{\rm fit1} \equiv \{q,\Gamma_{\rm peak},b,\alpha_1\},
\end{equation}
and
\begin{equation}
    \Theta_{\rm fit2} \equiv \{F_0,\Gamma_{\rm peak},\theta_{\rm jet},q,\alpha_2\}.
\end{equation}

The posterior $p(\Theta_{\rm fit} | D)$ is calculated from the likelihood $p( D|\Theta_{\rm fit} )$ and the prior $p(\Theta)$ via Bayes’ Theorem:
\begin{equation}
    p(\Theta_{\rm fit} | D) \propto p(\Theta_{\rm fit})p(D | \Theta_{\rm fit}).
    \label{equ:likelihood}
\end{equation}

The proportionality constant in equation (\ref{equ:likelihood}) serves the purpose of normalization and does not affect the MCMC analysis. When considering the data points $(t_i,F_i)$, which are assumed to be independent of each other and have Gaussian uncertainties $\sigma_i$, we can represent the likelihood as a conventional $\chi^2$ form:
\begin{align}
    p(D| \Theta_{\rm fit}) & \propto \rm exp(-\frac{1}{2}\chi^2),\\
    \chi^2 & = \sum_i\left(\cfrac{F_i - F_{\rm model}(t_i,\Theta_{\rm fit})}{\sigma_i}\right)^2,
\end{align}
where $F_{\rm model}(t_i, \Theta_{\rm fit})$ is the flux calculated from our model at the time $t_i$ with parameters $\Theta_{\rm fit}$. 

To restrict and determine the parameters of the fit based on prior information about the data and model, the prior distribution $p(\Theta_{\rm fit})$ is employed. The prior is assumed to be uniformly distributed within specific limits as provided in Table \ref{tab:values}. With the foundational theory in place, we proceed to initiate the MCMC analysis.

During the fitting, the high precision and abundance of data points before the jet break ($\sim$ 670 s) causes high weighting, which makes the data after the jet break unimportant. Consequently, a model with even no jet break could fit the data as well. To mitigate this effect, we randomly select 12 data points from the high-precision part (before 670 s), and keep all the data after the break, to form a new dataset for fitting. To minimize the uncertainty associated with random selection, we repeated this data sampling procedure 100 times.

The MCMC analysis is performed using the \textsl{emcee} \citep{emcee} python package. This algorithm employs an ensemble of ``walkers" moving simultaneously through parameter space instead of the standard single walker approach. Once the target distribution is defined, an initial state is chosen within the bounds of the parameters in $\Theta_{\rm fit}$, and the process of run$\_$mcmc is initiated. Sampling is performed for 3800 iterations after burn-in, generating approximately 550,000 samples of $p(\Theta_{\rm fit}| D)$. To visually present the outcomes of the MCMC sampling process, \textsl{corner} package \citep{corner} is utilized to plot the diagnostics. 

\subsection{The result}\label{subsec:result}

We focus on the afterglow data during the slow decay and steep decay phases. The light curve data is integrated in the energy range 0.3–5 TeV. To ensure reliable fitting results, we exclude the rising phase by selecting data points beyond the initial 50 seconds. After this procedure, MCMC simulation is employed to fit our models.

\begin{figure}
    \centering
    \includegraphics[width=1\linewidth]{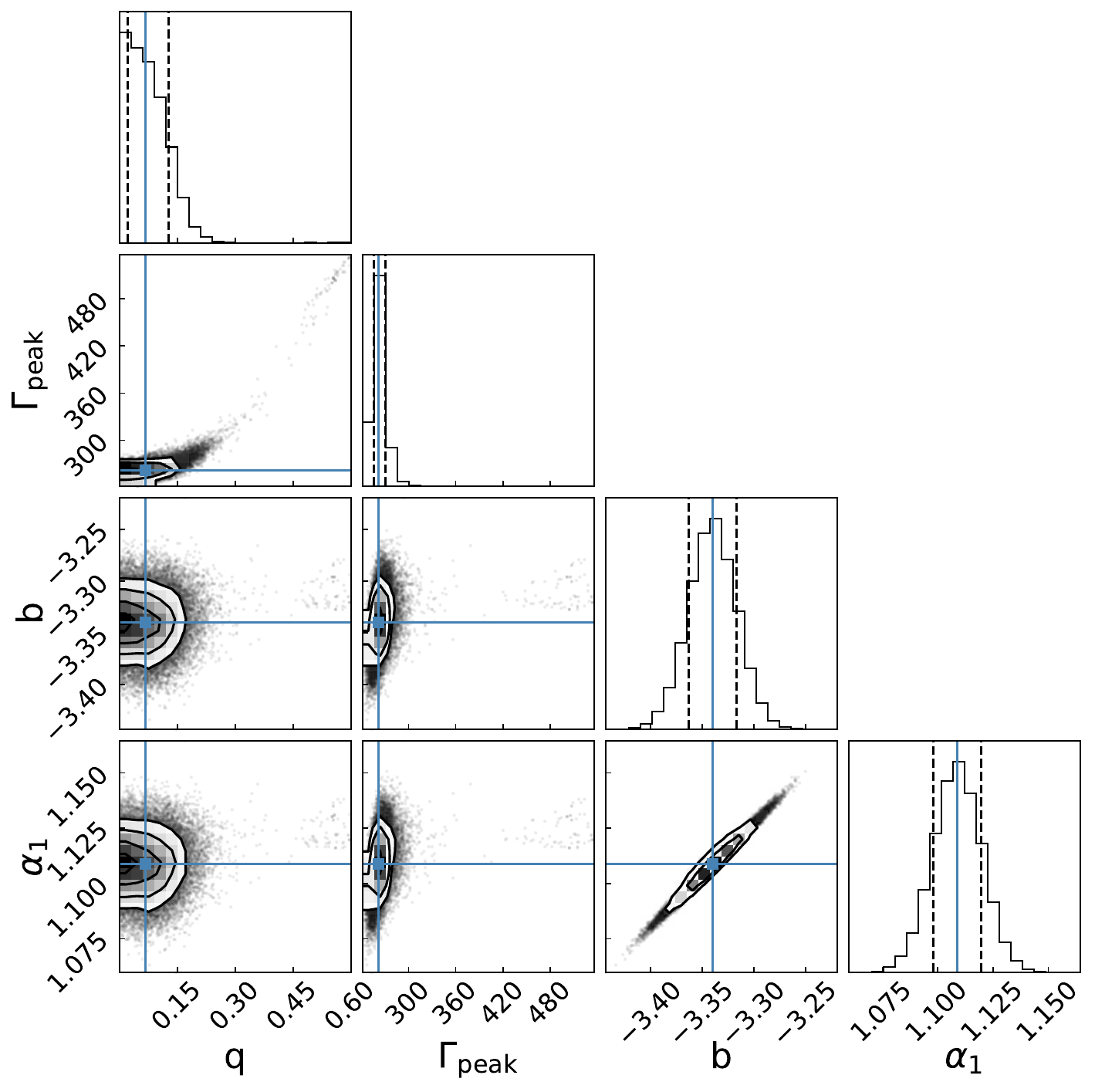}
    \caption{The fitting result of the geometry model (model 1). The best-fit values are highlighted in blue. The diagonals along the contours depict the marginalized probabilities for each parameter. The off-diagonal contours display the covariances between different parameter pairs.}
    \label{fig:mcmc-result}
\end{figure}

\begin{figure}
    \centering
    \includegraphics[width=1\linewidth]{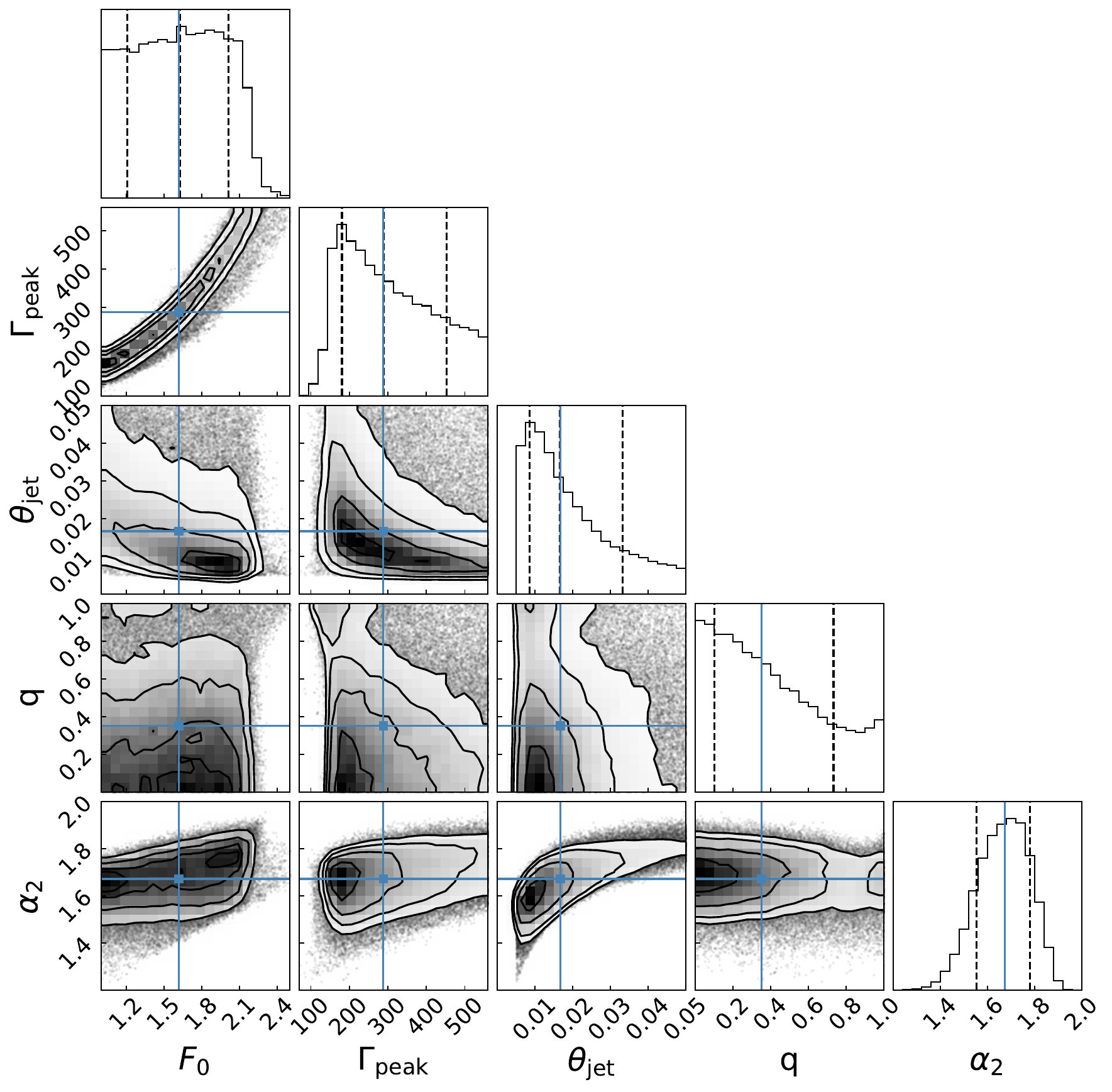}
    \caption{ The fitting result of the model considering high-latitude radiation (model 2). The best-fit values are highlighted in blue.  }  
    \label{fig:mcmc-result2}
\end{figure}

The MCMC analysis results in the generation of  contours, as depicted in Fig. \ref{fig:mcmc-result} and Fig. \ref{fig:mcmc-result2}. In these plots, each element corresponds to a histogram representing the distribution of an individual variable, while the correlations between variables are displayed within the contours. 

It becomes evident that the distribution of all four variables in Fig. \ref{fig:mcmc-result} closely resembles a Gaussian distribution. However, as shown in Fig. \ref{fig:mcmc-result2}, it is notable that the variables do not follow a Gaussian distribution for model 2. Still, they exhibit some form of concentration that indicating a set of best values.




By selecting the MAP (maximum posterior probability),  the best-fit values are determined. These values, along with their uncertainties, are presented in Table \ref{tab:values}.

\begin{figure}
    \centering
    \includegraphics[width=1\linewidth]{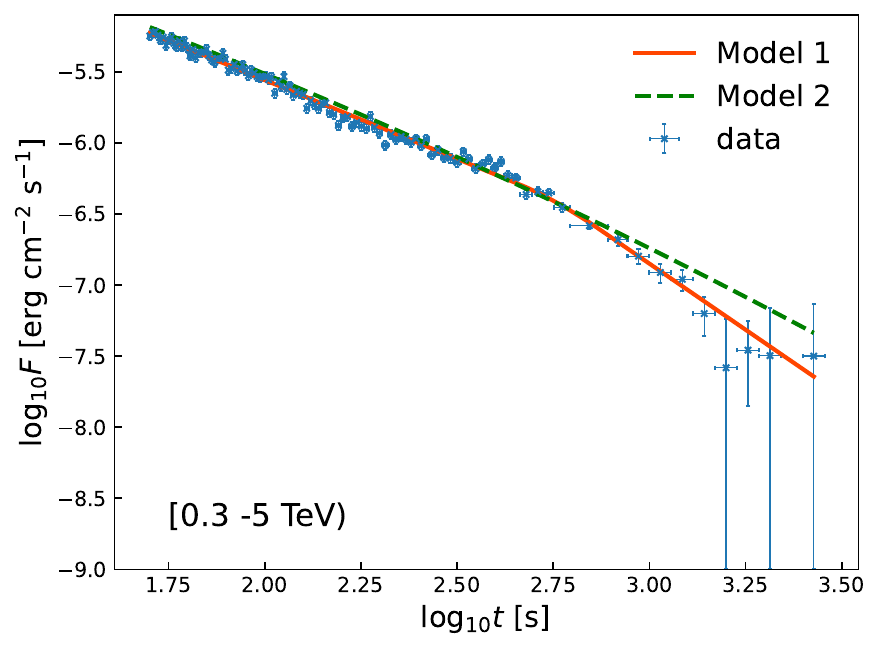}
    \caption{The fitting result of the light curve is presented with the best fitting values. The blue data points are the late TeV afterglow light curve of GRB 221009A, selected after the initial 50 seconds. The red solid curve is the best fitting result of model 1. The green dashed curve is the best fitting result of model 2 by fitting the sampled data as described in section \ref{sec:MCMC}.}
    \label{fig:curve-fit}
\end{figure}

Utilizing the determined best-fit values, we depict the fitting results in Fig. \ref{fig:curve-fit}. 
It displays the fitting results of the models and the TeV data after 50 seconds since the start of the TeV emission.
The consideration of high latitude radiation in the model fitting process is found to be less effective than excluding it, as high latitude radiation itself contributes to a smoothing effect on the curvature \citep[e.g.][]{2002ApJ...570L..61G}. The program successfully converges to stable parameter values, resulting in relatively small uncertainties. 
In our geometry model without considering high-latitude radiation, the outcome of the best-fit analysis indicates that the ratio q approximates 0.0673, while the viewing angle $\theta_{\rm obs}$ is estimated to be around 9.40$\times 10^{-4}$ rad, with an uncertainty of approximately 4.7$\times 10^{-4}$ rad. 
This result aligns with the assumption posited by \citep{2023Sci...380.1390L}, suggesting that the jet under consideration exhibits a highly focused core and is nearly aligned with our line of sight. 
However, upon accounting for high-latitude radiation, the results obtained through MCMC analysis appear to diverge from the previously obtained results. 
The best value of $\theta_{\rm jet}$ and $q$ after considering high-latitude radiation is 0.0169 rad and 0.3518, respectively. 
The value of $\theta_{\rm obs}$ is about 5.96$\times 10^{-3}$ rad, which is about 6 times greater than the previous results. However, in the context of comparing this value to the findings of other researchers, both results can be considered negligible or close to zero. 
This discrepancy between our models highlights the significant impact of model choice on the determination of $\theta_{\rm obs}$.

In Fig. \ref{fig:curve-fit}, one can see the model 1 (the simplified geometric model) fits the data even better than the model 2 (the model considering the high-latitude emission as well). We checked the dashed line by extending it to a much longer time, and it returns to the steeper slope same as the solid one. This indicates that the jet break has been smoothed in a much wider time range. 
It is reasonable that model 2 has a shallower light curve after the jet break compared with model 1 because the high-latitude emission can smear out the over simplified geometric model. 
However, it is a puzzle that the data matches the simplified model better (the red solid curves cross most of the data after break, while the green dashed curve does not). 
A structured jet cannot relieve this issue, as a top-hat jet is the sharpest structure. It might be a jet break overlapping with spectral frequency crossing, which makes the light curve steeping. However, the spectral index after the break is not softened as argued by \citet{2023Sci...380.1390L}.



\begin{table}
    \centering
    \begin{tabular}{p{0.92cm}|p{0.73cm}p{1.1cm}p{1.8cm}p{1.8cm}}
    \hline
        Model 1& Median & Bound & 1$\sigma$ uncertainty & 3$\sigma$ uncertainty\\
    \hline
        q & 0.0673 & [0, 1]   & (0.020, 0.127)  & (0.000, 0.255) \\
        $\Gamma_{\rm peak}$ & 261.8& (70,560)  & (255.5,270.1) & (245.9,304.5) \\
        b & -3.340&(-4,-2.5)  & (-3.363,-3.317) & (-3.406,-3.278) \\
        $\alpha_1$ & 1.109& (1.0,1.5)  & (1.098, 1.120) & (1.077, 1.138)   \\
    \hline
        Model 2& Median& Bound  & 1$\sigma$ Quantity & 3$\sigma$ Quantity\\
    \hline
        q & 0.3518 & [0, 1]  & (0.1041,0.7275) & (0.0017,0.9961) \\ 
        $\Gamma_{\rm peak}$ & 291.6 & (70,560)  & (183.6,453.7) & (111.7,557.8) \\ 
        $\theta_{\rm jet}$ & 0.0169 & (0.005,0.05)   & (0.0088,0.0334) & (0.0051,0.0496) \\ 
        $F_0$ & 1.6241 & (30,35)  & (1.2184,2.0098) & (1.0047,2.3414) \\ 
        $\alpha_2$ & 1.6757 & (1.11,2)  & (1.5549,1.7816) & (1.3558,1.8891) \\ 

    \hline
    \end{tabular}
    \caption{Best-fit values of $\Theta_{\rm fit1}$ and $\Theta_{\rm fit2}$ are listed. Bounds and their uncertainties are also shown. Model 1 only applies simple geometry, which is described in section \ref{subsec:without}. Model 2 is to consider the high-latitude radiation, which is described in section \ref{subsec:with}.}
    \label{tab:values}
\end{table}

\section{conclusion and discussion} \label{sec:con}

We developed two models to determine the viewing angle of GRB 221009A by fitting its TeV light curve measured by LHAASO, which exhibited a distinct jet break feature in the TeV band.
Both models treat the jet as having a top-hat angular distribution and assume the Lorentz factor decays as $\Gamma \propto t^{-3/8}$. 
Through the application of MCMC sampling, we derived optimal viewing angles of 9.4$\times 10^{-4}$ rad and 5.9$\times 10^{-3}$ rad. Both viewing angles are sufficiently small to be regarded as approaching zero, thus providing robust evidence supporting the on-axis assumption.

Our method has certain limitations. Firstly, we did not consider the structured jet model (either power law \citep{2001ApJ...552...72D, 2002MNRAS.332..945R}, or Gaussian \citep{2004ApJ...601L.119Z}). Given the complex nature of the environments in the TeV band, which require further discussion, it was challenging to construct a reliable model for fitting. The second issue is the determination of $\Gamma_{\rm peak}$. $\Gamma_{\rm peak}$ is assumed to be derived from the initial Lorentz factor $\Gamma_0$, a value already established by \cite{2023Sci...380.1390L}. However, the complex parameters of the environment hindered our ability to obtain a reliable $\Gamma_{\rm peak}$. Therefore, we opted to abandon the computation of an exact peak Lorentz factor in favor of fitting. In this context, the evolution of the jet within its environment is not accounted for. The third limitation in our study is that we solely focused on the TeV energy flux observed by LHAASO, neglecting multi-wavelength fitting. While this approach imposes certain restrictions on the scope of our study, it effectively shielded us from encountering challenging scenarios, such as data absence during jet breaks \citep[e.g.][]{2023ApJ...948L..12K}. As a result, we were able to draw a clear conclusion that the ejecta emitting TeV photons possesses a very small off-axis angle, essentially treatable as zero.

\begin{table}
    \centering
    \begin{tabular}{c|c|c|c}
    \hline
    \hline
        waveband & $\theta_{\rm jet}$ & $\theta_{\rm obs}$ &  \\
    \hline
    \hline
          radio to X-rays & $\ge$ 0.4 & $\lesssim$ 0.016  &    \cite{2023SciA....9I1405O} \\
        \hline
         radio to X-rays&$\approx $ 0.021  & $\sim $ 0.02  & \cite{2023MNRAS.524L..78G}   \\
        \hline
         2-8 keV X-ray &0.0262 & 0.0175    &   \cite{2023ApJ...946L..21N} \\
        \hline
         radio to TeV & 0.01    &   & \cite{2023MNRAS.522L..56S} \\
         & 0.1 &  &   \\
        \hline
        X-ray and optical  & 0.0245&   & \cite{2023ApJ...947...53R}   \\
        \hline
         0.3-10 keV X-ray & 0.0349 &   &   \cite{2023ApJ...946L..24W}  \\
        \hline
          X-ray and optical& $<$0.02 &    &\cite{2023ApJ...946L..28L} \\
        \hline
         Radio to GeV & 0.0286 &  &\cite{2023ApJ...946L..23L}\\
        \hline
        
          X-ray to  $\gamma$-ray& 0.0124 &      &\cite{2023arXiv230301203A}\\
        \hline
          X-ray to  $\gamma$-ray& 0.0127 &      &\cite{2024ApJ...962L...2Z}\\
        \hline
         TeV & 0.0105 & & \cite{2024JHEAp..41...42Z}\\ \hline
         X-ray and TeV& 0.0012 & & \cite{2024MNRAS.tmp..652D} \\
    \hline
         radio to TeV & 0.0234 & & \cite{2024ApJ...962..115R}\\
        & 0.0051 & & \\
        \hline
        TeV &0.0140 & 0.0009  & Model 1 (this work)\\
        & 0.0169 & 0.0060 & Model 2 (this work)\\
    \hline
    \hline
    \end{tabular}
    \caption{References related to jet opening angles and off-axis angles are listed. Blank spaces indicate cases where observing angles were not discussed. All these results are calibrated in units of radians.} 
    \label{tab:papper}
\end{table}

Noticing different authors have obtained different jet opening angle and viewing angle for this GRB, we have listed these angles from other literature in Table \ref{tab:papper} for a comparison. 
It is apparent that our results are significantly smaller compared to those reported by other researchers. 
This is very likely because of the data was used in different bands. For example, the ``jet break" is at around 670 s for TeV band \citep{2023Sci...380.1390L}, while the ``break" is 0.255 day for the ultraviolet band \citep{2023ApJ...946L..24W}, 0.98 day for the optical band \citep{2023ApJ...948L..12K}.
There is also a trend that opening angle is smaller, constrained by the lower bands (such as optical band and radio band). 
It might be that the jet opening angle is expanding while it penetrates in the environment medium, because of lateral expansion \citep{1999ApJ...525..737R}. 
It is also might be caused by the structured jet, e.g., two-component jet. The different viewing angles may correspond to the two components of the jet \citep{2023MNRAS.522L..56S}, i.e., the smaller opening angle component (with higher Lorentz factor) corresponds to the TeV emission, while the wider one corresponds to the lower band emission.

As illustrated in Fig. \ref{fig:curve-fit}, the rate of decay of jet bending after introducing the high-latitude radiation is noticeably slower compared to the scenario where high-latitude radiation is not considered. This suggests that the model incorporating high-latitude radiation may not be comprehensive, and there may be other influences leading to variations in the rate of jet breaking. We speculate that the jet emitting TeV photons may also have its own structure, or there are some spectral evolution overlapping at the jet breaking.

\section*{Acknowledgements}
We thank the helpful discussions with A. M. Chen, Weihua Lei,  Kai Wang, DuanYuan Gao, and the hospitality of Yao'an station of Purple Mountain Observatory.
The language was refined by ChatGPT.
This work is supported in part by the National SKA Program of China (2022SKA0130100), and by the science research grants from the China Manned Space Project with No. CMS-CSST-2021-B11.
The simulation was completed on the HPC Platform of Huazhong University of Science and Technology.

\section*{Data Availability}
The data used are publicly available at \cite{2023Sci...380.1390L} (Figs. 3 and 4), and can be downloaded from \url{https://www.nhepsdc.cn/resource/astro/lhaaso/paper.Science2023.adg9328/}. 



\bibliographystyle{mnras}
\bibliography{bib} 








\bsp	
\label{lastpage}
\end{document}